\documentclass[a4paper]{jpconf}
\usepackage{graphicx}
\usepackage{upgreek}
\usepackage{lineno}
\begin{document}
\title{Overview of Project 8 and Progress Towards Tritium Operation}

\author{Walter C. Pettus, on behalf of the Project 8 Collaboration}

\address{Center for Experimental Nuclear Physics and Astrophysics and Department of Physics, University of Washington, Seattle, WA 98195, USA}

\ead{pettus@uw.edu}

\begin{abstract}
Project 8 is a tritium endpoint neutrino mass experiment utilizing a phased program to achieve sensitivity to the range of neutrino masses allowed by the inverted mass hierarchy. The Cyclotron Radiation Emission Spectroscopy (CRES) technique is employed to measure the differential energy spectrum of decay electrons with high precision. We present an overview of the Project 8 experimental program, from first demonstration of the CRES technique to ultimate sensitivity with an atomic tritium source. We highlight recent advances in preparation for the first measurement of the continuous tritium spectrum with CRES.
\end{abstract}

\section{Introduction}

The existence of neutrino mass, which has been definitively proven by the detection of neutrino oscillations, is the only direct contradiction of the Standard Model of Particle Physics~\cite{Patrignani:2016xqp}.  Laboratory measurements of the beta spectrum endpoint are sensitive to the average mass of the electron flavor neutrino, $m_{\nu_e}$
\begin{equation}
\label{eqn:m_nu_e}
m_{\nu_e} = \sqrt{\sum_i{\left|U_{ei}\right|^2 m_i^2}} \quad ,
\end{equation}
where $U_{ei}$ are elements of the PMNS matrix and $m_i$ are masses of the neutrino mass eigenstates.  As $m_{\nu_e}$ is proportional to an incoherent sum over mass eigenstates, beta endpoint measurements are a direct and model independent probe of the absolute mass scale of neutrinos, complimentary to searches from cosmology and neutrinoless double beta decay~\cite{Drexlin:2013lha}.  Using tritium as the beta isotope, the Mainz and Troitsk experiments have achieved the greatest sensitivity among beta endpoint measurements, yielding a limit of $m_{\nu_e} < 2\,\mathrm{eV}$~\cite{Kraus:2004zw,Aseev:2011dq}.  The newest experiment based on the same MAC-E filter detector technology, KATRIN, aims to deliver an order-of-magnitude improvement in sensitivity~\cite{Angrik:2005ep} and is undergoing commissioning now.

Project 8 is a phased concept for a next-generation experiment with targeted sensitivity down to $m_{\nu_e} \sim 40$\,meV~\cite{Esfahani:2017dmu}.  The detection technique, based on the novel cyclotron radiation emission spectroscopy, represents a departure from the design of previous tritium spectrometers, enabling Project 8 to exhaust the remaining space allowed in the inverted mass hierarchy\footnote{The neutrino mixing angles and mass-squared differences measured by oscillation experiments constrain the terms of Eqn.~\ref{eqn:m_nu_e}, $m_{\nu_e} > 50\,\mathrm{meV}$ for the inverted mass hierarchy and $m_{\nu_e} > 9\,\mathrm{meV}$ for the normal mass hierarchy.}.

\section{Cyclotron Radiation Emission Spectroscopy}

Cyclotron Radiation Emission Spectroscopy (CRES) is a frequency-based detection technique allowing high precision spectroscopy of decay electrons~\cite{Monreal:2009za}.  Relativistic electrons in a magnetic field will emit cyclotron radiation at a frequency, $f_c$:
\begin{equation}
\label{eqn:f_c}
f_c = \frac{1}{2\pi} \cdot \frac{e B}{m + E_\mathrm{kin}/c^2} \quad ,
\end{equation}
related to their kinetic energy, $E_{kin}$, and the magnetic field, $B$; where $e$ is the electron charge, $m$ is the electron mass, and $c$ is the speed of light in vacuum.  For electrons near the 18.6\,keV tritium endpoint, the radiated Larmor power will be up to 1\,fW for a 1\,T field.

An energy resolution of 1\,eV can be achieved with this technique from a frequency resolution of 50\,kHz ($\Delta f / f = 2 \times 10^{-6}$), requiring a minimum observation time of 20\,$\upmu$s.  This necessitates the use of a no-work trap, such as a magnetic bottle, which confines the electron in the observation region.  Additionally, the gas density must be limited to maintain a sufficient mean free path despite electron-tritium inelastic scattering.

The characteristics of CRES make it well suited for performing tritium endpoint beta spectroscopy.  The high precision of the frequency measurement makes the energy resolution achievable, while the low radiated power can readily be amplified and recorded with modern microwave electronics.  There are no transport losses because the source is the detector and the gas transparency to its own cyclotron radiation places no intrinsic limitation on the volume scaling of the experiment.  CRES measures a differential spectrum with trigger rate limiting naturally obtained by setting the frequency window(s) of interest.  Finally, because the technique non-destructively monitors the electrons, systematics related to electron scattering and track reconstruction can be directly quantified from the data.

\section{Phase I}

\begin{figure}[b]
\includegraphics[width=23pc]{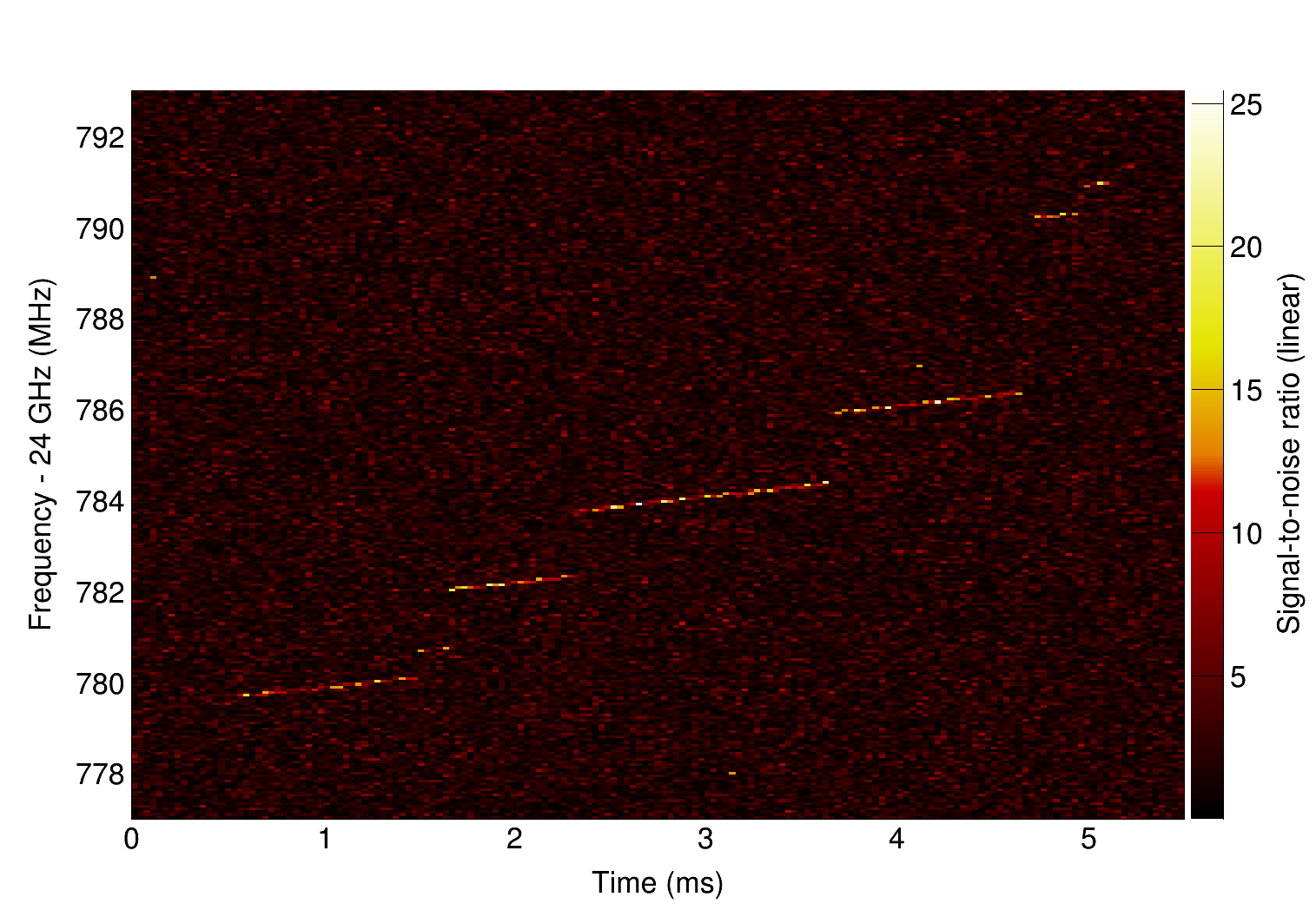}
\hspace{1pc}
\begin{minipage}[b]{13pc}\caption{\label{fig:event0}Spectrogram of the first electron recorded in Project 8 Phase I.  The track demonstrates the electron's slow loss of energy due to cyclotron emission and acute energy loss in a series of scatters with residual gas molecules.  The electron is born in the trap near 0.5\,ms and 780\,MHz, and scatters out of the trap after several milliseconds.}
\end{minipage}
\end{figure}

The initial phase of Project 8 was conceived as a demonstration of the CRES technique, providing the first detection and spectroscopy of single electrons via their relativistic cyclotron radiation.  This tabletop-scale experiment, located at the University of Washington, recorded its first electron events in 2014~\cite{Asner:2014cwa} (see Fig.~\ref{fig:event0}).

The source gas used is $^{83\mathrm{m}}$Kr, which produces a series of mono-energetic conversion electrons over the range 7\,--\,32\,keV; the K-shell line at 17.8\,keV is closely matched to the tritium endpoint of 18.6\,keV.  These conversion electrons provide sharp energy features for calibration over the full energy range of interest.  A $^{83}$Rb generator constantly releases $^{83\mathrm{m}}$Kr with its 1.83\,hr half-life into the gas system, while getters effectively remove non-noble residual gases.

The gas cell is constructed with kapton windows isolating a section of WR42 waveguide.  Coil windings around the waveguide in three locations provide a configurable magnetic trapping field.  The electron cyclotron radiation couples to the TE$_{10}$ mode of the waveguide and is enhanced by a pair of low-noise cryogenic amplifiers before passing out to the warm receiver chain for downmixing and digitizing.  A warm-bore NMR magnet provides the main 1\,T field, but also limits the apparatus with its 50\,mm bore diameter.

\begin{figure}[h]
\includegraphics[width=20pc]{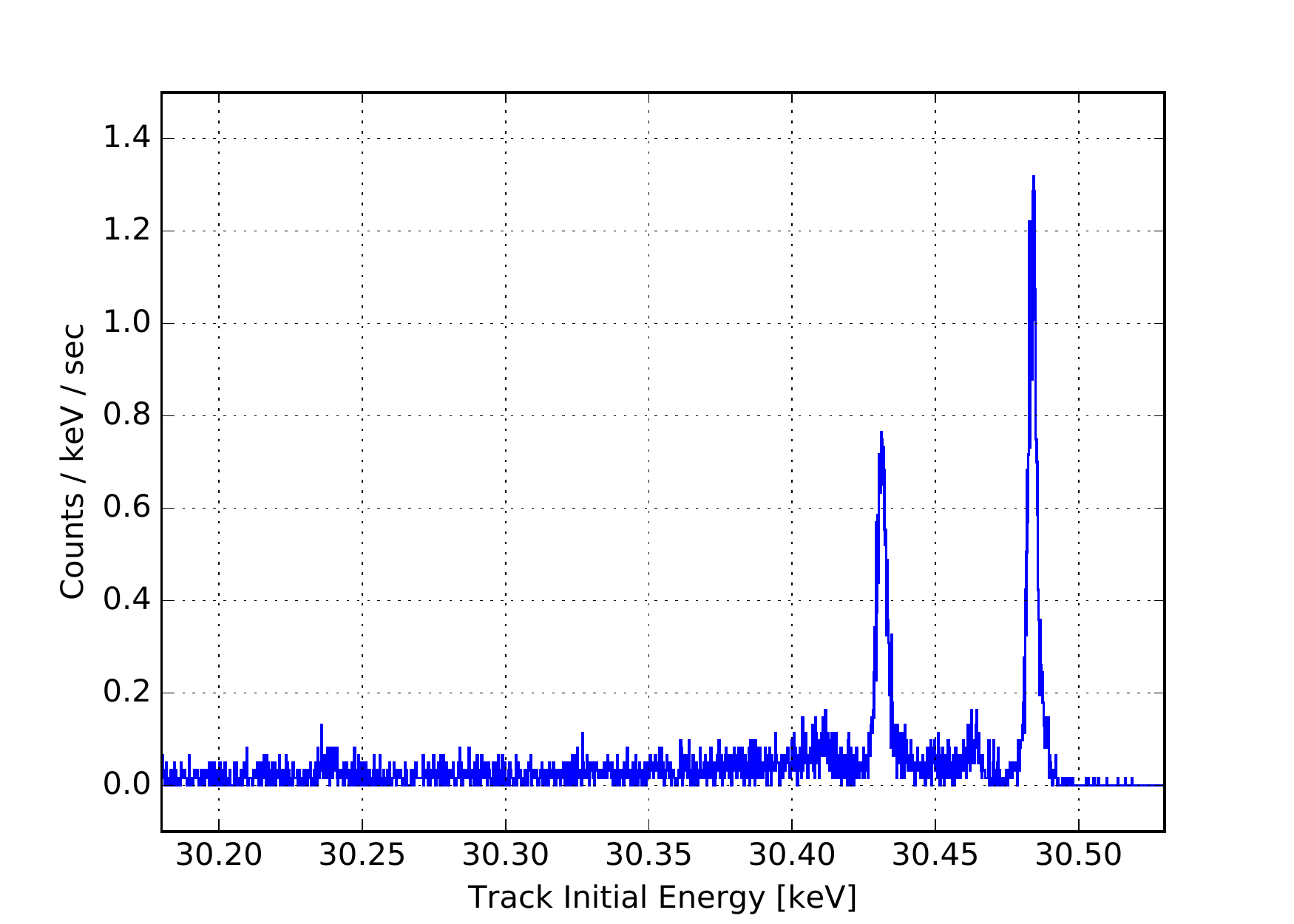}
\begin{minipage}[b]{18pc}\caption{\label{fig:30keV}Energy spectrum near the 30.4\,keV L-shell doublet in optimized trap geometry.  A cut on track start time with 70\% selection efficiency has been applied to remove poorly reconstructed events and noise triggers. The doublet lines, with separation of 52.8\,eV, are clearly resolved by the CRES technique.  The observed resolution of 3.3\,eV FWHM is approaching the natural linewidths of 1.84 and 1.4\,eV in this doublet.}
\end{minipage}
\end{figure}

Following the initial demonstration of CRES, the trapping field configuration was optimized to improve the measured energy resolution.  A trap with a long uniform bottom region exhibits resolution approaching the natural linewidth of the electron capture lines (see Fig.~\ref{fig:30keV})~\cite{Esfahani:2017dmu}.  The corresponding raw spectrograms also have sideband tracks related to the axial motion of the electrons in the trap, allowing an event reconstruction sensitive to pitch angle in the trap~\cite{Guigue:2017}.

\section{Phase II}

The second phase of Project 8 will make the first measurement of the continuous tritium spectrum using the CRES technique.  It builds on the successes of Phase I by reusing the same magnet and amplifiers, but with a redesigned insert featuring tritium compatibility and an upgraded analog signal path.  Phase II will also be a platform for testing alternative data acquisition (DAQ) systems extendable for future phases with more channels.

\begin{figure}[t]
\centering
\includegraphics[width=32pc]{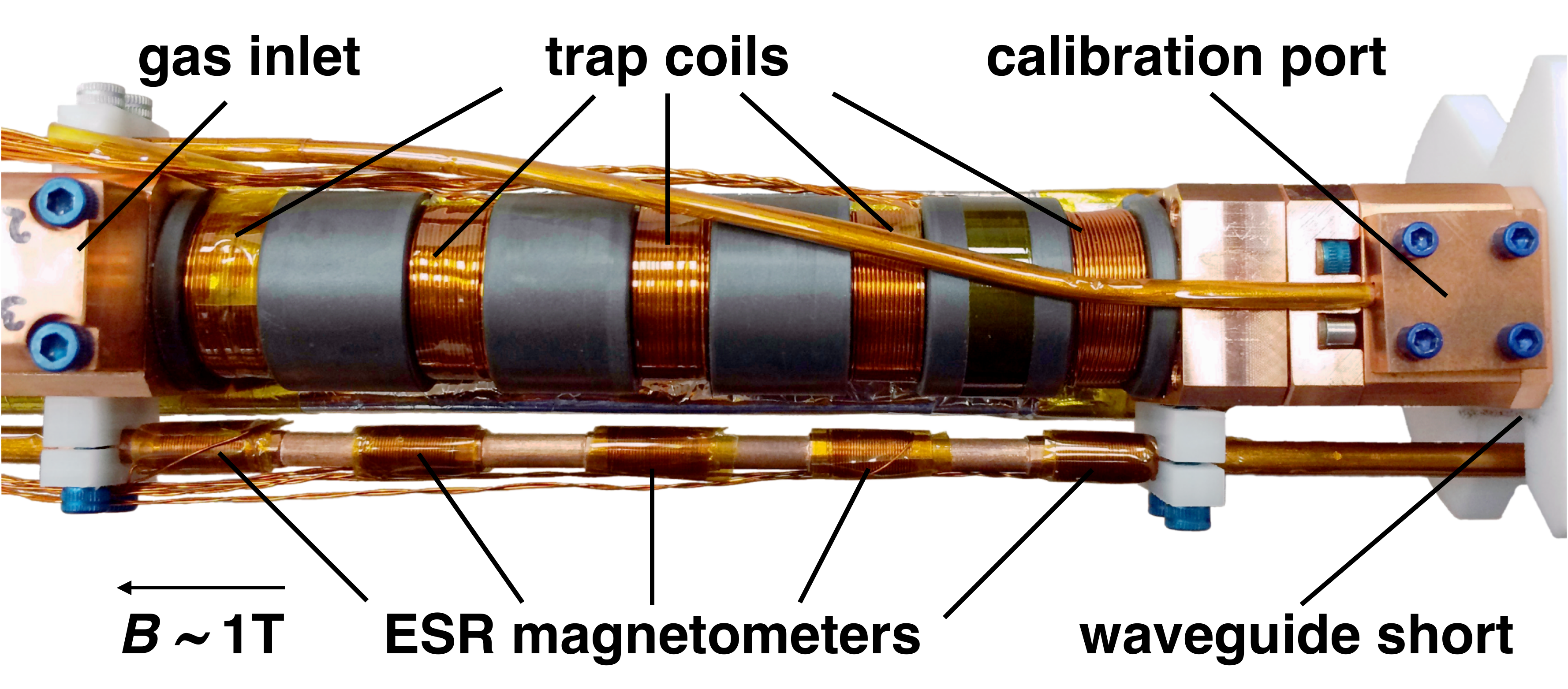}
\caption{\label{fig:phase2_cell}Phase II tritium-compatible gas cell.  The cell features an enlarged volume and more trapping coils relative to Phase I, as well as ESR magnetometers to monitor the magnetic field \textit{in situ}.  The cell is rotated for insertion into the magnet so that the B-field is aligned upwards in the vertical direction.}
\end{figure}

The Phase II gas cell is built on a circular waveguide matched to the circularly polarized cyclotron radiation.  The circular waveguide also accommodates a significant increase in gas volume due to its larger cross section and increased length.  Tritium compatibility is established through the use of CaF$_2$ windows sealed with indium gaskets.  The longer cell is equipped with five current coils to provide greater tunability in the trapping field.  The field contribution from each coil can be individually monitored by the five off-axis electron spin resonance (ESR) magnetometers (see Fig.~\ref{fig:phase2_cell}).

Above the gas cell, a circulator with cold termination improves the signal-to-noise of the system by both flattening the noise spectrum and decreasing its overall level.  A quarter-wave plate is also inserted to transition between the cell and the rectangular polarization acceptance of the amplifiers.

\begin{figure}[t]
\centering
\includegraphics[width=36pc]{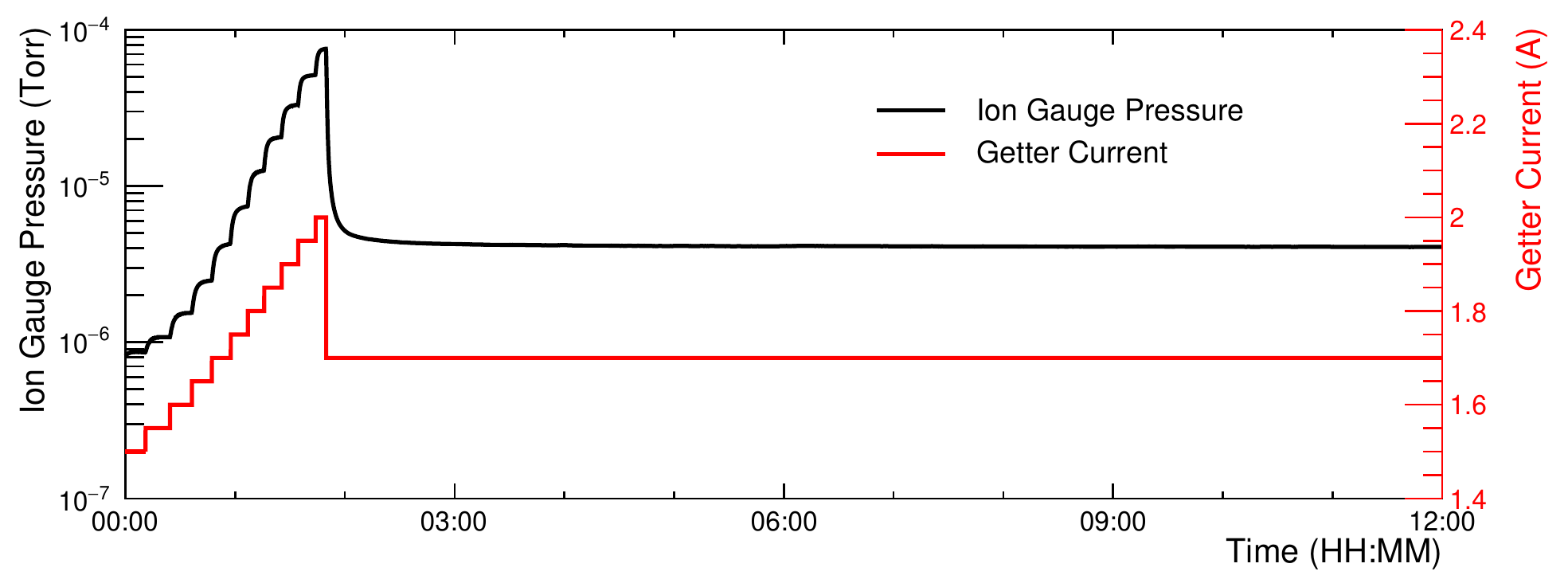}
\caption{\label{fig:getter}Tritium gas system pressure regulation with getter during deuterium commissioning test.  For the first two hours, the getter current is regularly increased in steps to demonstrate the tunability of the pressure setpoint.  After 02:00, the getter current is held constant to demonstrate pressure stability.}
\end{figure}

A new gas system capable of running both tritium and krypton has recently been commissioned.  A non-evaporable getter stores the tritium and maintains the equilibrium gas pressure at a tunable density around the nominal operating point of $10^{11}$\,cm$^{-3}$ (see Fig.~\ref{fig:getter}).  A series of remote-controlled valves and automatic shutoffs ensures safe operation and control of the gas system.  The safety review of this gas system has been newly completed in summer 2017, clearing Phase II for first injection of tritium into the cell.

\section{Phase III}

The third phase of Project 8 will demonstrate scalability of the CRES technique by leaving the confines of a waveguide and instrumenting a 200\,cm$^3$ gas volume to collect the free-space cyclotron radiation.  This scale of experiment targets a neutrino mass limit of 2\,eV, comparable to the current experimental limits~\cite{Kraus:2004zw,Aseev:2011dq}.

\begin{figure}[t]
\centering
\includegraphics[width=12pc]{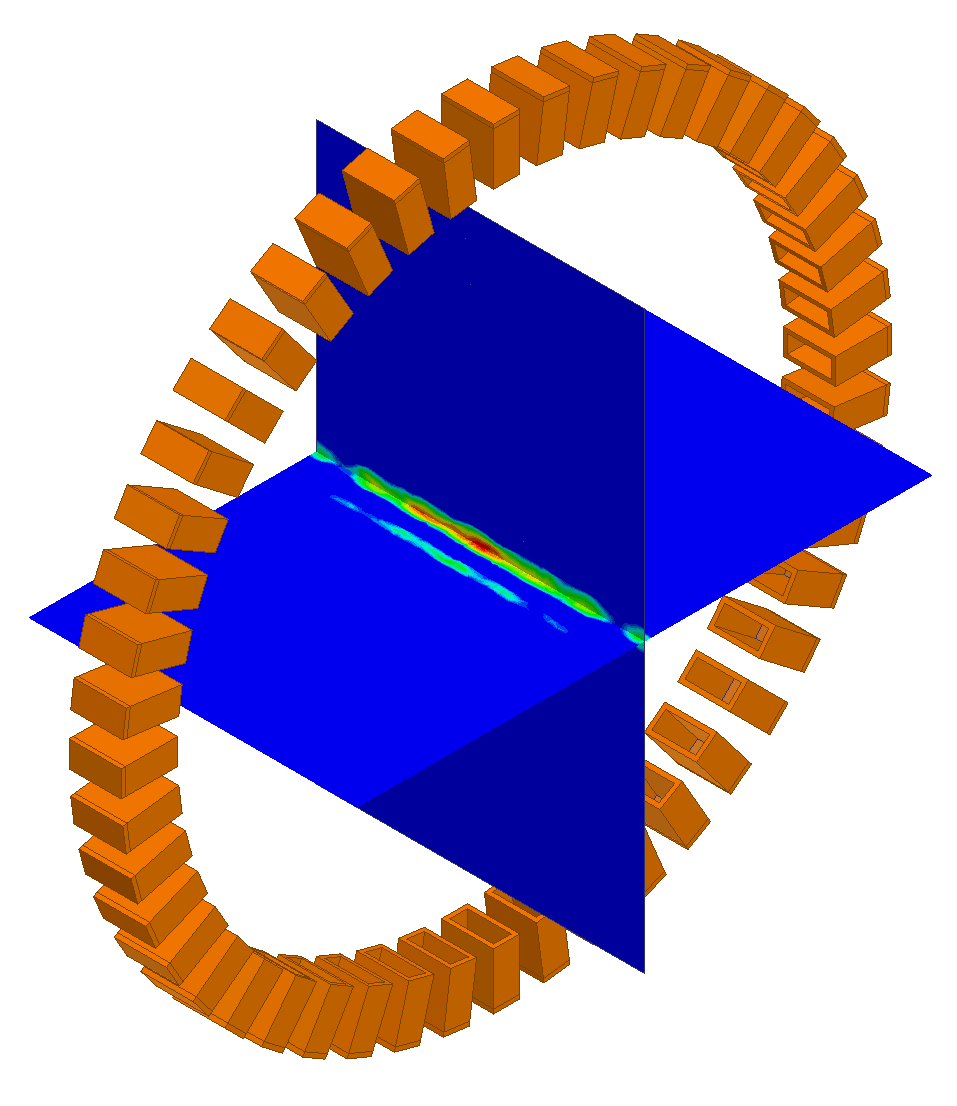}
\hspace{3pc}
\includegraphics[width=18pc]{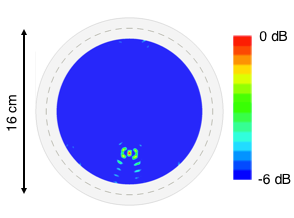}
\caption{\label{fig:phase3}Simulation of digital beam-forming based on 48-element ring array of antennas.  {\it Left:} focus of array along central $z$-axis.  {\it Right:} focus of array in $xy$-plane.}
\end{figure}

Collecting the signal will require an array of antennas surrounding the gas volume (see Fig.~\ref{fig:phase3}).  Digital beam-forming with appropriate phase delays can achieve localization of the electron signal for fiducialization and noise suppression.  Working groups have formed to begin implementing this detector concept, with particular focus on optimizing signal collection, triggering and event reconstruction from multiple DAQ channels, trapping geometry, and design of the physics insert.  This geometry will be accommodated within a typical surplussed hospital MRI magnet.

\section{Phase IV}

The fourth phase of Project 8 targets a neutrino mass sensitivity down to 40\,meV by implementing an atomic tritium source.  This will overcome the irreducible systematic limitation of molecular tritium sources at the 100\,meV sensitivity level incurred from the width of the molecular excitation final state distribution~\cite{Bodine:2015sma}.  The active volume required to acquire the necessary statistics will be $\sim$100\,m$^3$, and the limiting systematic is projected to be a magnetic field uncertainty of $\Delta B/B \sim 10^{-7}$.  A working group has formed to plan demonstrators to prepare for the challenges of producing, cooling, and trapping the required atomic tritium.

\section{Conclusion}

Project 8 presents a next-generation neutrino mass experiment concept with sensitivity down to $m_{\nu_e} \sim 40$\,meV, completely covering the allowed range of the inverted hierarchy.  Results from Phase I provide the first demonstration of the CRES technique and exhibit eV-level energy resolution.  Phase II has recently commissioned a tritium gas system and will shortly begin the first CRES measurement of the tritium spectrum.  In parallel, an active R\&D program is being pursued to enable significant scaling in volume towards Phase III and to prepare for the challenges of operating with atomic tritium in Phase IV.


\end{document}